\documentclass[iop]{emulateapj}
\usepackage{graphicx}
\usepackage{epstopdf}

\shorttitle{The kpc-scale star formation law at z=4}
\shortauthors{Hodge et al.}

\slugcomment{Accepted to ApJ Letters}

\begin{document}

\title{The kiloparsec-scale star formation law at redshift 4: wide-spread, highly efficient star formation in the dust-obscured starburst galaxy GN20}

\author{J. A. Hodge\altaffilmark{1,2,3}}
\altaffiltext{1}{National Radio Astronomy Observatory, 520 Edgemont Road, Charlottesville, VA, 22903, USA}
\altaffiltext{2}{Max-Planck-Institut f\"{u}r Astronomie, K\"{o}nigstuhl 17, 69117 Heidelberg, Germany}
\email{jhodge@nrao.edu}
\altaffiltext{3}{Jansky Fellow}

\author{D. Riechers\altaffilmark{4}}
\altaffiltext{4}{Department of Astronomy, Cornell University, Ithaca, New York, 14853, USA}

\author{R. Decarli\altaffilmark{2}}

\author{F. Walter\altaffilmark{2}}

\author{C. L. Carilli\altaffilmark{5,6}}
\altaffiltext{5}{National Radio Astronomy Observatory, P.O. Box 0, Socorro, NM 87801-0387, USA}
\altaffiltext{6}{Astrophysics Group, Cavendish Laboratory, JJ Thomson Avenue, Cambridge CB3 0HE, UK}

\author{E. Daddi\altaffilmark{7}}
\altaffiltext{7}{CEA, Laboratoire AIM-CNRS-Universit\'{e} Paris Diderot, Irfu/SAp, Orme des Merisiers, F-91191 Gif-sur-Yvette, France}

\author{H. Dannerbauer\altaffilmark{8}}
\altaffiltext{8}{Universit\"{a}t Wien, Institut f\"{u}r Astrophysik, T\"{u}rkenschanzstra\ss e 17, 1180 Wien, Austria}

\begin{abstract}
We present high-resolution observations of the 880\,$\mu$m  (rest-frame FIR) continuum emission in the z$=$4.05 submillimeter galaxy GN20 from the IRAM Plateau de Bure Interferometer (PdBI). These data resolve the obscured star formation in this unlensed galaxy on scales of 0.3$^{\prime\prime}$$\times$0.2$^{\prime\prime}$ ($\sim$2.1$\times$1.3\,kpc). The observations reveal a bright (16$\pm$1\,mJy) dusty starburst centered on the cold molecular gas reservoir and showing a bar-like extension along the major axis.  The striking anti-correlation with the HST/WFC3 imaging suggests that the copious dust surrounding the starburst heavily obscures the rest-frame UV/optical emission. A comparison with 1.2\,mm PdBI continuum data reveals no evidence for variations in the dust properties across the source within the uncertainties, consistent with extended star formation, and the peak star formation rate surface density (119$\pm$8\,M$_{\sun}$\,yr$^{-1}$\,kpc$^{-2}$) implies that the star formation in GN20 remains sub-Eddington on scales down to 3\,kpc$^2$. We find that the star formation efficiency is highest in the central regions of GN20, leading to a resolved star formation law with a power law slope of $\Sigma_{\rm SFR}$\,$\sim$\,$\Sigma_{\rm H_2}^{\rm 2.1\pm1.0}$, and that GN20 lies above the sequence of normal star-forming disks, implying that the dispersion in the star formation law is not due solely to morphology or choice of conversion factor.  These data extend previous evidence for a fixed star formation efficiency per free-fall time to include the star-forming medium on $\sim$kpc-scales in a galaxy 12\,Gyr ago.

\noindent\textit{Key words:} galaxies: evolution $-$ galaxies: formation $-$ galaxies: high-redshift $-$ galaxies: ISM $-$ galaxies: star formation

\end{abstract}
\section{INTRODUCTION}
\label{Intro}

One of the most commonly used diagnostics in studies of galaxy formation is the relation between gas surface density ($\Sigma_{\rm gas}$) and star formation rate (SFR) surface density ($\Sigma_{\rm SFR}$), which describes the relative efficiency at which gas is transformed into stars in different environments \citep[e.g.][]{1959ApJ...129..243S, 1998ARA&A..36..189K}. Recent studies indicate 
 that this star formation (SF) law (or `Kennicutt-Schmidt' law) is molecular  \citep[e.g.,][]{2008AJ....136.2846B, 2008AJ....136.2782L, 2011AJ....142...37S}, 
and that it does not evolve with redshift \citep[e.g.,][]{2010ApJ...714L.118D, 2010MNRAS.407.2091G}. There is evidence, however, for two different SF regimes on the ${\rm \Sigma}_{\rm SFR}$--${\rm \Sigma}_{\rm H_2}$ plane (`main sequence galaxies' versus `starbursts'; \citealt{2010ApJ...714L.118D, 2010MNRAS.407.2091G}, but c.f. \citealt{2012MNRAS.421.3127N}).

Due to the difficulty of resolving the cold molecular gas and obscured star-forming regions in distant star-forming galaxies, studies of the high-redshift SF law have historically been limited to unresolved detections. However, it is clear that resolved observations
are necessary to test whether the gas and SFR surface densities on sub-galactic scales are significantly different than those implied by global averages,
as well as
whether the most intensely star-forming galaxies at high-redshift -- which appear to have $\Sigma_{\rm SFR}$ that are consistent with theories of so-called ``maximum starbursts''  \citep[e.g.][]{1999ApJ...517..103E, 2006ApJ...640..228T} -- are approaching the Eddington limit for dust 
in radiation pressure-supported disks \citep{2003JKAS...36..167S, 2005ApJ...630..167T} on $\sim$kpc, or even galaxy-wide scales \citep[e.g.,][]{2009Natur.457..699W, 2013Natur.496..329R}. 
While such resolved studies have become possible in the last few years, they are still limited to either $z$$<$2 \citep{2013A&A...553A.130F} or gravitationally-lensed galaxies \citep{2010Natur.464..733S, 2012ApJ...752....2D, 2013ApJ...765....6S, 2014ApJ...783...59R}.

To this end, we here report high-resolution rest-frame far-infrared (FIR) imaging of GN20 \citep{2006MNRAS.370.1185P, 2009ApJ...694.1517D}, 
a $z$$=$4.05 submillimeter galaxy \citep[SMG;][]{2014arXiv1402.1456C}.
GN20 has the deepest, highest-resolution cold molecular gas imaging currently available for any unlensed high-$z$ galaxy, revealing an extended ($\sim$14$\pm$4 kpc) gas reservoir resolved on scales of $\sim$1.3 kpc \citep[hereafter H12]{2012ApJ...760...11H}.
The observations presented here provide nearly matched-resolution imaging of the dust emission,
allowing us to investigate the SF efficiency (SFE) and SF law on $\sim$kpc scales 12\,Gyr ago.
Where applicable we assume a concordance, flat $\Lambda$CDM cosmology \citep[H$_0$=71\,km\,s$^{-1}$\,Mpc$^{-1}$, $\Omega_{\Lambda}$=0.73, $\Omega_{M}$=0.27;][] 
{2007ApJS..170..377S}.  

\begin{figure*}[t!]
\centering
\includegraphics[scale=0.65]{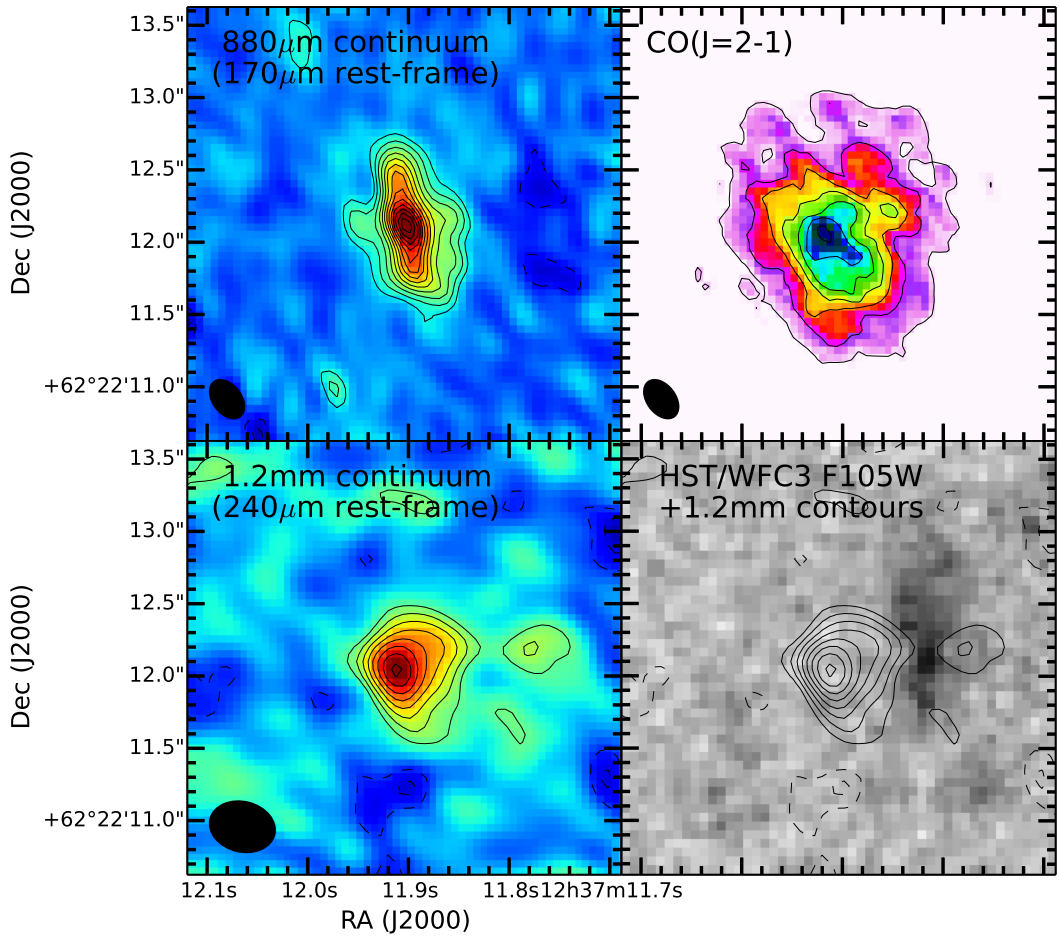}
\vspace{-2.1cm}
\caption{Multiwavelength imaging of the z$=$4.05 SMG GN20.
\textit{\textbf{ Top left:}} 880$\mu$m PdBI image (0.3$''$$\times$0.2$''$ resolution). Contours start at $\pm$2$\sigma$ in steps of 1$\sigma$$=$0.25mJy\,beam$^{-1}$ (corresponding to a rest-frame brightness temperature of T$_{B}$$=$0.22K). The measured peak flux density of 3.6mJy implies  
T$_{B,peak}$$=$3.2K. 
\textit{\textbf{Top right:}} VLA CO(2--1) 0th moment map (H12) at the same resolution as the 880$\mu$m data. 
\textit{\textbf{Bottom left:}} 1.2mm PdBI image (0.46$''$$\times$0.35$''$ resolution). Contours start at $\pm$2$\sigma$ in steps of 1$\sigma$$=$0.25\,mJy\,beam$^{-1}$.
\textit{\textbf{Bottom right:}} HST/WFC3 F105W image from the CANDELS survey \citep{2011ApJS..197...35G} 
and 1.2mm contours. 
 }
\label{multiwav}
\end{figure*}

\section{OBSERVATIONS \& DATA REDUCTION}
\label{obs}

\subsection{PdBI 880$\mu$m}
\label{880obs}

The 880\,$\mu$m (170\,$\mu$m rest--frame) observations of GN20 ($\alpha$(J2000)$=$12h37m11.920s, $\delta$(J2000)$=$62$^{\circ}$22$'$12.0$''$) 
were carried out in two tracks on 2013 December 4 (C-configuration track) and 2014 March 7 (A-configuration track) using six antennas. 
The observations used the PdBI's Band 4, tuned to 340\,GHz (880\,$\mu$m), along with the WideX correlator (3.6\,GHz bandwidth).
The receiver was operating in the upper side band. The nearby radio quasars B1044+719 
and B1418+546 were used for pointing, amplitude, and phase calibration, and the flux calibrators were 3C279 and MWC349 (A-track) and 3C84 and LKH$\alpha$101 (C-track). We estimate the flux calibration to be good to within 20\%.

The IRAM \textsc{gildas} package was used for data reduction and analysis.
The calibration process included two iterations of phase-only self-calibration, and one iteration of amplitude$+$phase self-calibration.
After flagging, there were 
5.7\,h/4.3\,h on-source in the C/A-tracks (6-antenna equivalent).
The final map (C$+$A tracks; Figure~\ref{multiwav}) has 0.05$^{\prime\prime}$ pixels and was created using the \textsc{clean} algorithm with robust weighting and a tight clean box on the source (cleaning down to 2$\sigma$).
The image has a synthesized beam of 0.3$^{\prime\prime}$$\times$0.2$^{\prime\prime}$ and an rms of 0.25\,mJy\,beam$^{-1}$. As the source lies at the phase center of the 14.8$^{\prime\prime}$ primary beam, the map was not corrected for primary beam attenuation.

\begin{figure}[]
\centering
\includegraphics[scale=0.77]{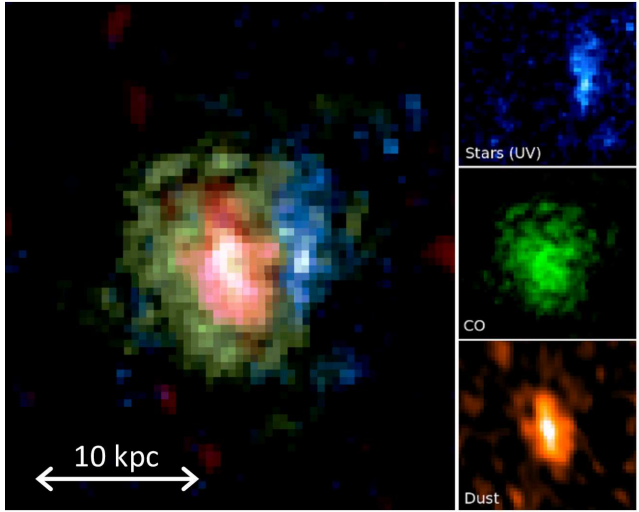}
\caption{False-color image of GN20, showing the HST/WFC3 F105W emission (tracing young stars; blue), the VLA CO(2--1) emission (tracing cold molecular gas; green), and the PdBI 880$\mu$m emission (tracing dust-obscured SF; red). 
}
\label{falsecolor}
\vspace{1mm}
\end{figure}


\subsection{PdBI 1.2mm}
\label{1mmobs}

The 1.2\,mm (240$\mu$m rest--frame) PdBI observations of GN20 (20.1$"$ primary beam) were carried out in two tracks on 2010 February 1 and 7 using six antennas in the A-configuration.
The 250\,GHz observations used the previous generation correlator, providing a total bandwidth of 1.0\,GHz (dual polarization). 
The quasars B1044+719, B1418+546, and B1300+580 were used for 
pointing, amplitude and phase calibration. 
Several standard calibrators (MWC349, Titan, 3C273, 3C279, 3C345, 3C454.3, B0234+285, B0923+392, B1055+018, B1749+096) were observed for flux and bandpass calibration, yielding $\sim$15\% calibration accuracy.

The data were mapped using \textsc{clean} with natural weighting, resulting in a synthesized beam of 0.46$''$$\times$0.35$''$ (Figure~\ref{multiwav}). 
Weighting schemes that would lead to higher spatial resolution were discarded due to high noise. The final rms noise is 
0.25\,mJy beam$^{-1}$ over the full bandpass.

\subsection{VLA CO(2--1)}
\label{COobs}

The CO(2--1) data on GN20 were obtained with the Karl G. Jansky Very Large Array (VLA) and presented in \citet{2011ApJ...739L..33C} and H12. 
The 0th moment map used here (Figure~\ref{multiwav}) was created by taking the native resolution (0.19$^{\prime\prime}$/1.3\,kpc) data cube and convolving it to the resolution of the 880\,$\mu$m data (0.3$^{\prime\prime}$$\times$0.2$^{\prime\prime}$), applying the same mask as used in H12.


\section{RESULTS}
\label{results}

\subsection{Spatially-resolved rest-frame FIR emission}
\label{FIRresults}

We have detected and spatially-resolved the 880\,$\mu$m (170\,$\mu$m rest--frame) continuum emission from GN20 (Figure~\ref{multiwav}). We measure a peak flux density of 3.6\,mJy\,beam$^{-1}$ at $\alpha$(J2000)$=$12h37m11.90s, $\delta$(J2000)$=$62$^{\circ}$22$'$12.10$''$, positionally coincident with the peak of the CO emission within 0.1$^{\prime\prime}$, 
and suggesting that the most intense SF occurs in the region with the highest concentration of cold gas. 
The integrated flux density (16$\pm$1\,mJy) is consistent 
with the integrated flux density measured in the C-configuration data alone (18\,mJy; beam 0.88$''$$\times$0.74$''$), indicating that the higher-resolution data are not resolving out a significant amount of emission. These measurements are also consistent with the single dish measurement for GN20 \citep[S$_{\rm 850}$$=$20.3$\pm$2.1\,mJy;][]{2006MNRAS.370.1185P}. Note that despite its large flux density, the evidence suggests that GN20 is not lensed \citep{2010ApJ...714.1407C}.

The full size of GN20's molecular gas reservoir is 14$\pm$4\,kpc (H12). From the new 880\,$\mu$m continuum data, we find that the dust-obscured SF is also extended, with a bar-like extension along the major axis of the galaxy. 
We measure a deconvolved source size of (0.76$''$$\pm$0.06$''$)$\times$(0.33$''$$\pm$0.03$''$) (FWHM, equivalent to $\sim$5.3$\times$2.3\,kpc), implying a physical size of $\sim$10\,kpc out to 10\% of the peak flux density (along its major axis). This suggests that the dust-obscured SF in this galaxy is almost as extended as the gas reservoir -- a conclusion which is consistent with the spatially-resolved excitation structure (H12).  

While the FIR emission is coincident with the CO emission, both the FIR and CO emission are 
offset by $\sim$0.6$^{\prime\prime}$ (4\,kpc) from the peak of the rest-frame UV emission as traced by the HST/WFC3 F105W image (Figure~\ref{falsecolor}).
The striking anti-correlation suggests that the copious dust surrounding the starburst heavily obscures the UV/optical light produced by newly-formed stars and keeps it from escaping the galaxy in all but one small region several kpc from the nuclear burst.

We have also detected and resolved GN20 in the 1.2\,mm (240$\mu$m rest--frame) continuum observations (Figure~\ref{multiwav}). 
We measure an integrated flux density of 5.7$\pm$1.6\,mJy, which is $\sim$2/3 of the flux measured
with the MAMBO-2 bolometer at 11$''$ resolution at the same wavelength \citep[9.3$\pm$0.9\,mJy;][]{2008MNRAS.389.1489G}. 
As the 1.2\,mm data were taken entirely in the 
extended configuration, this could indicate that some spatially-extended flux on $\gtrsim$1$''$ scales is resolved out. The FWHM source size of 
(0.77$''$$\pm$0.17$''$)$\times$(0.49$''$$\pm$0.17$''$)
measured from the 1.2\,mm observations
 is consistent with that measured from the 880\,$\mu$m data. 

\begin{figure}[]
\centering
\includegraphics[scale=0.46]{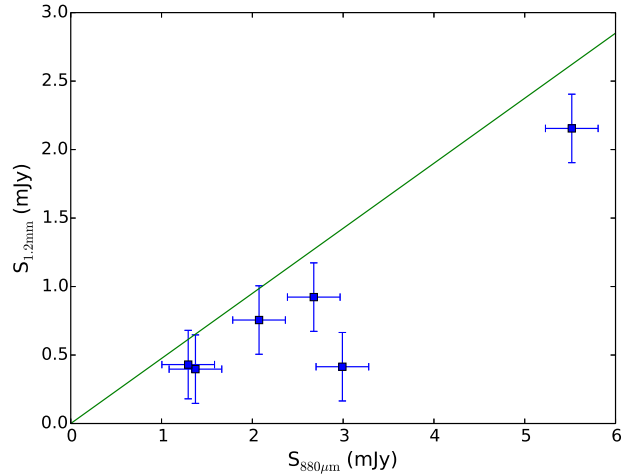}
\caption{S$_{\rm 880\mu m}$ vs. S$_{\rm 1.2mm}$ for
0.4$''$ ($\sim$2.8\,kpc) resolution elements.
The solid line represents the ratio expected from the best-fit model to GN20's IR SED \citep{2014A&A...569A..98T}. 
}
\label{dustcompare}
\vspace{0mm}
\end{figure}

\begin{figure*}[]
\centering
\includegraphics[scale=0.49]{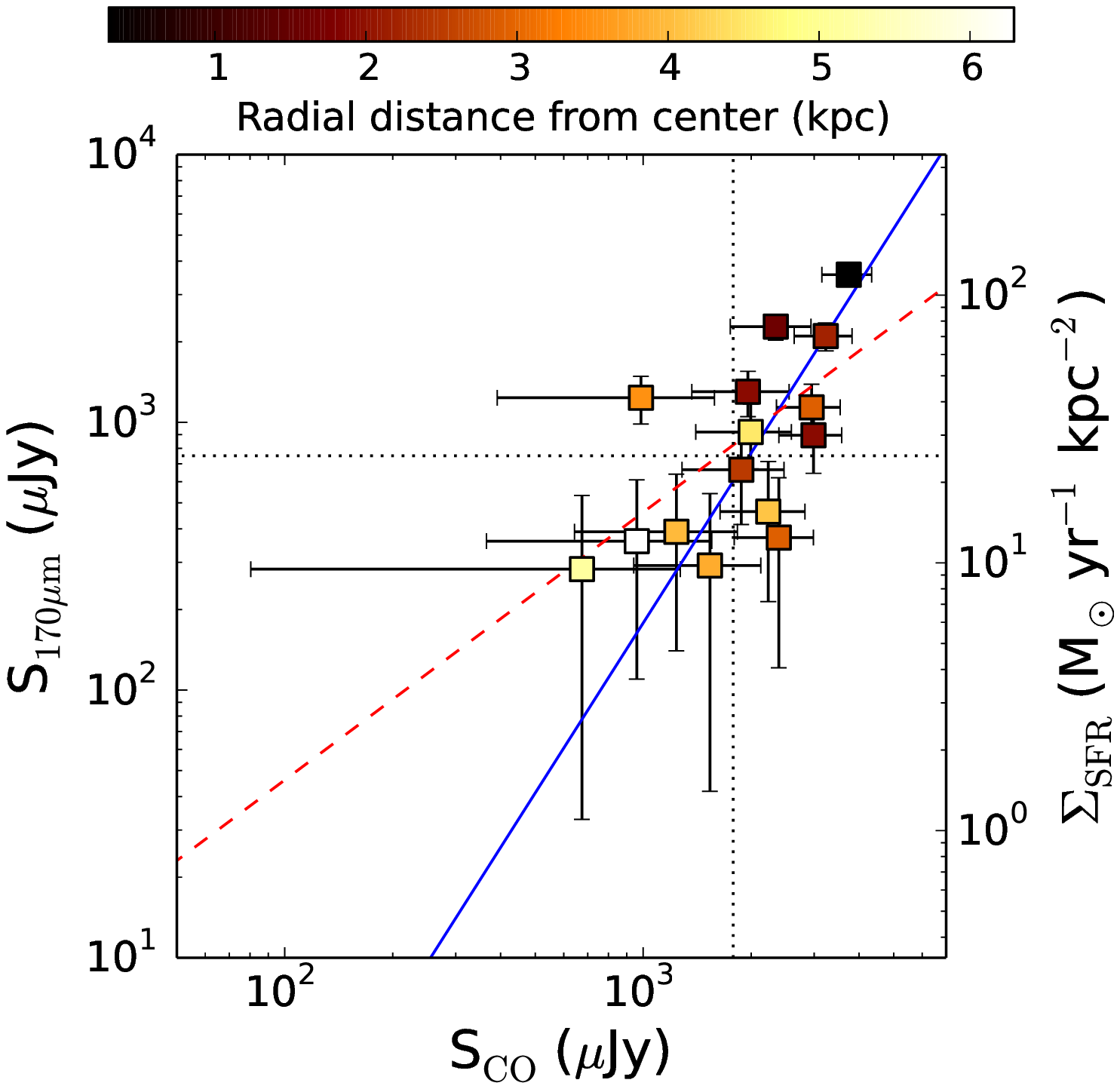}
\hfil
\hspace{-4mm}
\includegraphics[scale=0.49]{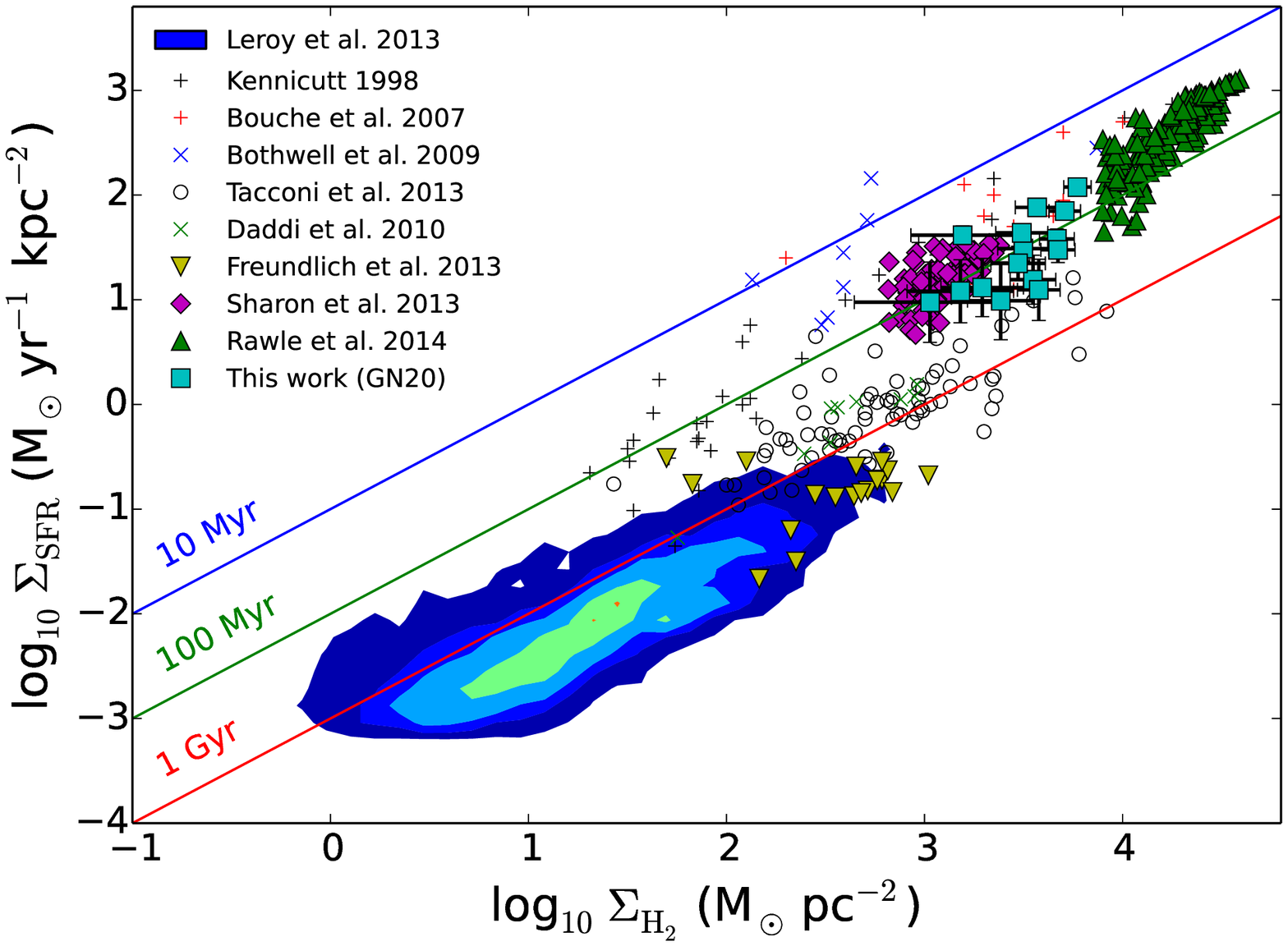}
\caption{The relation between molecular gas and SF in GN20. 
\textit{\textbf{Left:}} Rest-frame 170\,$\mu$m flux density vs. 
CO(2--1) flux density, 
color-coded by distance from the galactic center.
The right-hand axis shows $\Sigma_{\rm SFR}$ (see Section~\ref{SFRresults}). The solid blue line shows a power law fit to the data, including points below 3$\sigma$ (indicated by black dotted lines), and the red dashed line shows a slope of unity 
(indicating constant SFE).
\textit{\textbf{Right:}} $\Sigma_{\rm SFR}$ vs. $\Sigma_{\rm H_2}$ in GN20 and other local and high-redshift sources from the literature. 
Contours indicate the density of 
local galaxies \citep[taken from][]{2013AJ....146...19L}. 
Unfilled data points show unresolved measurements, including local ULIRGs \citep[black plus-signs;][]{1998ApJ...498..541K}; 
z$\sim$0.5 disk galaxies and z$\sim$1.5 BzK galaxies \citep[green crosses;][]{2010ApJ...714L.118D}; 
z$\sim$1-3 color-selected galaxies \citep[black circles;][]{2010Natur.463..781T}; 
and SMGs as red plus-signs \citep{2007ApJ...671..303B} and blue crosses \citep{2010MNRAS.405..219B}. 
Filled symbols show resolved measurements, including z$\sim$1.2 massive star-forming galaxies \citep[yellow upside-down triangles;][]{2013A&A...553A.130F}; two strongly-lensed SMGs as magenta diamonds \citep{2013ApJ...765....6S} and green triangles \citep{2014ApJ...783...59R}; and this work on GN20 (cyan squares).
The solid lines indicate constant gas depletion timescales of 10\,Myr (blue), 100\,Myr (green), and 1\,Gyr (red). 
} 
\label{KSlaw}
\end{figure*}

\subsection{Dust continuum slope}
\label{dustslope}

We used our dust imaging to examine possible deviations from a constant 880\,$\mu$m/1.2\,mm ratio, which could suggest variations in the dust continuum slope across the galaxy. 
A gradient in the slope could indicate a gradient in the dust temperature and/or optical depth -- both of which may be expected for dense nuclear starbursts. 
We convolved the 880\,$\mu$m image to the resolution of the 1.2\,mm image, resampled both maps to have 0.4$''$ pixels ($\sim$2.8\,kpc, the approximate size of the resolution element), and blanked all pixels except those with flux densities $>$1$\sigma$ in both maps in a central contiguous region. The resulting flux densities are plotted in Figure~\ref{dustcompare}, 
where
the solid line represents the ratio expected from the best-fit \citet{2007ApJ...657..810D} dust model to the IR SED
-- see \citeauthor{2014A&A...569A..98T} (2014, hereafter T14) for further details. 
In general, we find 1.2\,mm flux densities that are $\sim$20-25\% lower than expected, 
providing further evidence that the lack of short baseline information may have resolved out some extended emission.\footnote{Another possibility is that the 880$\mu$m flux density is boosted by the presence of H$_{2}$0(3$_{03}$-2$_{12}$) emission ($\nu_{rest}$$=$1716.769633\,GHz), but taking its 3$\sigma$ upper limit (given its non-detection), and assuming the CO(2--1) linewidth (730\,km\,s$^{-1}$), we limit its contribution to $\lesssim$3.5\% of the measured continuum flux.}  
The remaining scatter, 
however, is not statistically significant. 
We conclude that the data are consistent with a constant ratio, with no evidence for a variation in the dust continuum slope across the galaxy. 
This finding is consistent with the apparently extended nature of the SF in GN20, and it supports the use of a constant 
$\Sigma_{\rm SFR}$/S$_{880\mu m}$ conversion factor
in the analysis that follows.

\subsection{SFR surface densities}
\label{SFRresults}

GN20 has an infrared luminosity of log($L_{\rm IR}$/L$_{\odot}$)$=$13.27$\pm$0.02, implying a total SFR of 1860$\pm$90\,M$_{\sun}$\,yr$^{-1}$ (assuming a Chabrier IMF; T14).
This is consistent with an analysis of the PAH emission \citep{2014ApJ...786...31R}, indicating no significant AGN contamination to $L_{\rm IR}$.
Using the extent of the 880\,$\mu$m emission, the average $\Sigma_{\rm SFR}$ is $\sim$100\,M$_{\sun}$\,yr$^{-1}$\,kpc$^{-2}$. 
To determine how $\Sigma_{\rm SFR}$ varies on $\sim$kpc scales, we resampled the 880\,$\mu$m map onto a grid of 0.25$''$/1.75\,kpc pixels
and used the best-fit \citet{2007ApJ...657..810D} dust model to the IR SED (T14) to determine 
a conversion factor between the 880\,$\mu$m flux densities and IR luminosities of 1.006($\pm$0.045)$\times$10$^{12}$\,L$_{\odot}$\,mJy$^{-1}$. 
The uncertainty derives from the uncertainty on the (model-derived) $L_{\rm IR}$, which was quantified by T14 using a series of MC simulations. 
We then divided by the beam area and scaled based on the total SFR/$L_{\rm IR}$ for GN20.
We find values in the $\sim$tens of M$_{\sun}$\,yr$^{-1}$\,kpc$^{-2}$, peaking at 119$\pm$8\,M$_{\sun}$\,yr$^{-1}$\,kpc$^{-2}$ in the galaxy's center.

\subsection{The resolved SF law at z$=$4}
\label{KSresults}

In order to study the spatially-resolved SF law in GN20, we 
resampled the CO(2--1) map onto the pixel grid described in Section~\ref{SFRresults}. 
The observed flux densities from the CO and 880\,$\mu$m maps are plotted 
in Figure~\ref{KSlaw} (left). The ratio between FIR and CO emission (indicating SFE, and thus gas depletion timescale) is highest in the central regions of the galaxy, which have the highest gas columns. The SFE tends to decrease further from the galactic center, with an overall variation of 
$\sim$8 across the regions where most of the SF takes place. 
A function of the form $\Sigma_{\rm 880\mu m}$ $\sim$ ($\Sigma_{\rm CO})^{\rm N}$
was fit to the data using the IDL routine \textsc{linmix$\_$err} \citep{2007ApJ...665.1489K}, a linear regression estimator that uses a Bayesian approach and considers measurement errors in two variables. 
The result of the fit gives a slope of N$=$2.1$\pm$1.0.
This slope assumes a constant $\Sigma_{\rm SFR}$/S$_{\rm 880\mu m}$ ratio, as supported by our analysis in Section~\ref{dustslope}. 
To simulate the effect of a variable dust temperature across the source, we re-ran the \citeauthor{2007ApJ...657..810D} models with a wide range of ionization parameters (U$_{\rm min}$$=$10--100), resulting in $\Sigma_{\rm SFR}$ increasing/decreasing by less than a factor of two. 
Given the radial trend in SFE, this would increase the vertical scatter in the data, steepening the slope. 

Figure~\ref{KSlaw} (right) shows the GN20 data on the $\Sigma_{\rm SFR}$-$\Sigma_{\rm H_2}$ plane. 
In order to convert the CO(2--1) flux densities to molecular gas masses, we have assumed the standard relationships from \citet{2013ARA&A..51..105C} and a CO-to-H$_2$ conversion factor of $\alpha_{\rm CO}$$=$1.1\,M$_{\sun}$(K\,km\,s$^{-1}$\,pc$^2$)$^{-1}$. This value -- which is slightly higher than the typically-adopted ULIRG value -- 
was determined in H12 by 
using the dynamical mass (derived by modeling the CO(2--1) dynamics) to constrain the gas content\footnote{As $\alpha_{\rm CO}$ was dynamically-constrained directly from CO(2--1), there is no need to assume a CO(2--1)/CO(1--0) excitation ratio.}. 
The mean depletion time we measure for GN20 is 130\,Myr, with a range of 40--300\,Myr. 
A compilation of literature studies is shown for comparison, where all SFR measurements have been converted to a common \citet{2003ApJ...586L.133C} IMF. The gas masses were calculated using the CO-to-H$_2$ conversion factors assumed in the respective studies: $\alpha_{\rm CO}$$=$0.8 for the unresolved ULIRGs/SMGs \citep[including Helium;][]{1998ApJ...498..541K, 2007ApJ...671..303B, 2010MNRAS.405..219B}; 
$\alpha_{\rm CO}$$=$3.6 for the z$\sim$0.5 disk galaxies and BzKs \citep{2010ApJ...714L.118D}; 
$\alpha_{\rm CO}$$=$4.35 for the local spirals and color-selected galaxies \citep{2010Natur.463..781T, 2013AJ....146...19L, 2013A&A...553A.130F};
and 
$\alpha_{\rm CO}$$=$4.6 \citep{2013ApJ...765....6S} and 
$\alpha_{\rm CO}$$=$0.7 \citep{2014ApJ...783...59R}
for the two strongly-lensed resolved SMGs\footnote{Units same as above.}$^{,}$\footnote{Note that the \citeauthor{2014ApJ...783...59R} and \citeauthor{2013ApJ...765....6S} data are oversampled -- see their papers for details.}.

\begin{figure}[]
\centering
\includegraphics[scale=0.47]{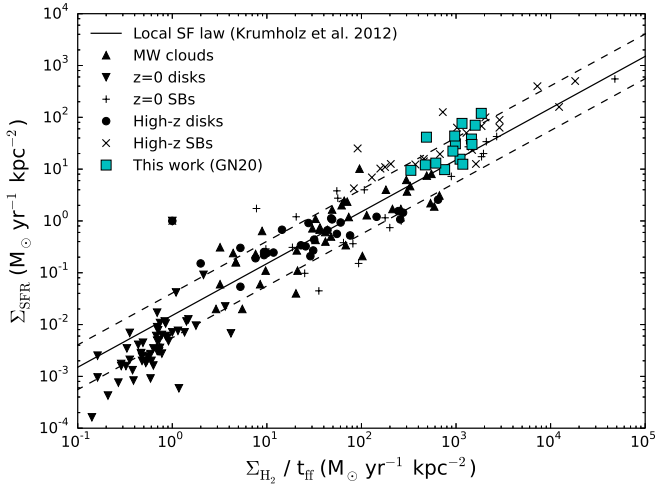}
\caption{$\Sigma_{\rm SFR}$ vs. $\Sigma_{\rm H_2}$/t$_{\rm ff}$ for GN20 (cyan squares) and a variety of comparison sources 
\citep{2012ApJ...745...69K}. The black solid (dashed) lines show the local volumetric SF law (and scatter) from \citet{2012ApJ...745...69K}, which assumes their best-fit value of $\epsilon_{\rm ff}$$=$0.015.}
\label{KS_tff}
\end{figure}

Figure~\ref{KS_tff} shows $\Sigma_{\rm SFR}$ versus $\Sigma_{\rm H_2}$, where the latter is divided by free-fall time (t$_{\rm ff}$). 
The comparison data are from \citet{2012ApJ...745...69K}. 
To calculate t$_{\rm ff}$ for GN20, we used equation~8 from \citeauthor{2012ApJ...745...69K}, which is appropriate for high surface density galaxies in the Toomre regime, and we have assumed the Toomre parameter Q$=$1, the dimensionless constant $\phi_{P}$$=$3 \citep{2005ApJ...630..250K}, the logarithmic index of the rotation curve $\beta$$=$0 (flat rotation curve; H12), and the angular velocity of galactic rotation $\Omega$$=$0.175\,Myr$^{-1}$ (calculated at the half-light radius). For GN20, this gives values of the SFE per free-fall time ($\epsilon_{\rm ff}$; i.e., the fraction of gas converted into stars per free-fall time) ranging from $\epsilon_{\rm ff}$$=$0.01-0.18.  
Also shown is the local volumetric SF law from \citet{2012ApJ...745...69K}, evaluated at their best-fit value of $\epsilon_{\rm ff}$$=$0.015.

\section{DISCUSSION}
\label{discussion}

While it is evident that SMGs host some of the largest starbursts in the known Universe ($\sim$10$^3$\,M$_{\odot}$\,yr$^{-1}$), this knowledge is largely based on integrated measurements and/or indirect SFR tracers. The latter imply that SMGs have typical SFR densities of $\sim$80\,M$_{\odot}$\,yr$^{-1}$\,kpc$^{-2}$ \citep{2006ApJ...640..228T}, 
well below the $\sim$1000\,M$_{\odot}$\,yr$^{-1}$\,kpc$^{-2}$ limit based on a theoretical description of the (dust--opacity) Eddington--limited SF of a radiation pressure--supported starburst on kpc scales \citep{2003JKAS...36..167S, 2005ApJ...630..167T},
though exceptions do exist \citep[e.g., $\sim$600\,M$_{\odot}$\,yr$^{-1}$\,kpc$^{-2}$ in the z$=$6.3 SMG HFLS3;][]{2013Natur.496..329R}.  
Indeed, the highest--resolution existing submillimeter observations on the lensed z$\sim$2.3 ``Eyelash" galaxy show that the star-forming regions are undergoing maximal starbursts \citep{2010Natur.464..733S}. 
Previous 0.8$^{\prime\prime}$ 890\,$\mu$m imaging \citep{2008ApJ...688...59Y} of GN20 -- which is intrinsically much more luminous than the Eyelash -- indicated that it might be forming stars close to or at its Eddington limit, although subsequent CO imaging \citep{2010ApJ...714.1407C} revealed the large gas disk and, along with a revised total SFR \citep{2009ApJ...694.1517D}, implied a sub-Eddington average value. Still, it was unknown whether the SF in GN20 was more compact than the cold gas reservoir traced by CO(2--1), or whether the indirectly-derived average value served to mask extreme differences in $\Sigma_{\rm SFR}$ between clump/inter-clump regions. 
Our high-resolution 880\,$\mu$m imaging shows that the SF in GN20 remains sub-Eddington on scales down to a few kpc$^2$.

An examination of the resolved SF law in this unlensed z$\sim$4 galaxy 
produced a power-law slope of N$=$2.1$\pm$1.0. While the significant uncertainty on the slope means that it may be only barely steeper than the local linear relation \citep[e.g.,][]{2013AJ....146...19L}, other studies also find evidence for a steepening in gas-rich/high density environments \citep[i.e.\,$\geq$200\,M$_{\sun}$\,pc$^{-2}$;][]{1987ApJ...319..730S, 2004ApJ...606..271G, 2005ApJ...623..826R},
including the centers of nearby galaxies \citep[e.g.,][]{2008ApJ...681L..77N, 2013AJ....146...19L}.
A slope of N$\sim$1.5 may be expected if SFR is proportional to gas mass divided by free-fall time \citep[assuming a constant scale height; e.g.,][]{1977MNRAS.178....1M, 2009ApJ...699..850K}. 
Other models of feedback-regulated SF \citep[e.g.][]{2011ApJ...731...41O, 2013MNRAS.433.1970F} give a quadratic slope -- closer to that seen here, though it is impossible to differentiate between models given the current observational uncertainties.

When compared with other studies
on the ${\rm \Sigma}_{\rm SFR}$--${\rm \Sigma}_{\rm H_2}$ plane, 
we find that the GN20 data lie above the sequence traced out by `normal' star-forming disk galaxies. 
This is consistent with its classification as a starburst (T14) and, given that the CO(2--1) data show evidence for a smoothly rotating gas disk (H12), it implies that the presence of an ordered velocity field is not a clear indication of the presence or absence of intense starburst activity. 
The approximate alignment of GN20 and the two (resolved) lensed SMGs is also notable. 
Given that the conversion factors specifically-derived for the individual cases range from ULIRG-like \citep{2014ApJ...783...59R} to slightly above ULIRG-like (GN20) to Galactic \citep{2013ApJ...765....6S}, 
this suggests that 
the dispersion in the SF law is apparently not due solely to choice of $\alpha_{\rm CO}$. 
On the contrary, the GN20 results suggest that the dispersion in the SF law is real, and that SMGs can have very high star formation efficiencies on small scales. 

If one assumes a universal SF law where the SFR volume density is simply a function of the molecular gas volume density and local t$_{\rm ff}$ \citep{1959ApJ...129..243S, 1998ApJ...498..541K, 2005ApJ...630..250K, 2008AJ....136.2782L}, then the complexity in the $\Sigma_{\rm SFR}$-$\Sigma_{\rm H_2}$ plane should be removed by accounting for t$_{\rm ff}$. 
Studies substituting the global dynamical timescale (t$_{\rm dyn}$) for t$_{\rm ff}$
in disk galaxies and mergers at low/high-redshift \citep{2010ApJ...714L.118D, 2010MNRAS.407.2091G} 
provided preliminary support for this theory, 
as t$_{\rm dyn}$ is proportional to t$_{\rm ff}$ for galactic scales in high surface density galaxies \citep{2005ApJ...630..250K, 2008AJ....136.2782L, 2010MNRAS.407.2091G, 2012ApJ...745...69K}. 
More recently, \citet{2012ApJ...745...69K} 
presented a complete theoretical framework for a local volumetric SF law, including a model for the local (star-forming cloud) t$_{\rm ff}$, 
which allowed them to  
describe the SF on scales from entire high-redshift galaxies down to individual molecular clouds with a fixed value of $\epsilon_{\rm ff}$$=$0.015.
Figure~\ref{KS_tff} demonstrates that our resolved imaging of GN20 is also consistent with 
this model,
extending the evidence for a fixed SFE per free-fall time to include the star-forming medium on $\sim$kpc-scales in a galaxy 12\,Gyr ago.

\acknowledgements
We thank John Hibbard, Mark Krumholz, Adam Leroy, Kartik Sheth, and an anonymous referee for helpful comments/discussions, 
and Roberto Neri, Sabine K\"{o}nig, and Melanie Krips for help with data analysis.  
IRAM is supported by INSU/CNRS (France), MPG (Germany) and IGN (Spain).  


\end{document}